\documentclass{ifacconf}

\usepackage{graphicx}      
\usepackage{natbib}        

\usepackage{amsmath,amssymb,amsfonts}
\usepackage{algpseudocode}
\usepackage{mathtools}
\usepackage{mathrsfs}
\usepackage{array}
\usepackage{xcolor}
\usepackage[absolute]{textpos} 

\newcommand{\R}{\mathbb{R}}
\newcommand{\mL}{\mathcal{L}}

\newcommand{\trace}{\mathop{\mathrm{tr}}}

\DeclareMathOperator*{\blkdiag}{blockdiag}

\begin{document}
\begin{frontmatter}

\title{On Data-based Nash Equilibria in LQ Nonzero-sum Differential Games\thanksref{footnoteinfo}} 

\thanks[footnoteinfo]{This work has received funding from the European Research Council (ERC) under the European Union’s Horizon 2020 research and innovation programme (grant agreement No 948679).}

\author[First]{Victor G. Lopez,} 
\author[First]{Matthias A. Müller}

\address[First]{Leibniz University Hannover, Institute of Automatic Control, 30167 Hannover, Germany (e-mail: \{ lopez, mueller \} @irt.uni-hannover.de).}

\begin{abstract}                
This paper considers data-based solutions of linear-quadratic nonzero-sum differential games. Two cases are considered. First, the deterministic game is solved and Nash equilibrium strategies are obtained by using persistently excited data from the multiagent system. Then, a stochastic formulation of the game is considered, where each agent measures a different noisy output signal and state observers must be designed for each player. It is shown that the proposed data-based solutions of these games are equivalent to known model-based procedures. The resulting data-based solutions are validated in a numerical experiment.
\end{abstract}


\end{frontmatter}


\begin{textblock*}{\textwidth}(1.5cm,29.3cm)
	\small{
		© 2026 the authors. This work has been accepted to IFAC for publication under a Creative Commons Licence CC-BY-NC-ND.\\
	}
\end{textblock*}

\section{Introduction}

Nonzero-sum (NZS) differential games model the interaction of several systems or processes, regarded as players, that influence each other while trying to achieve their individual goals. The theory of differential games has been used to analyze optimal behavior in multiagent dynamical systems for applications in diverse areas of study including engineering, economics, sociology, among many others \citep{BasarOls1999}. In a game among rational agents, the most important solution concept occurs when each player uses its best policy with respect to the best policy of the other players. These strategies are said to form a Nash equilibrium. Differential games have been studied for many decades \citep{StarrHo1969,BasarOls1999,Engwerda2005} and the conditions for the existence of Nash equilibria are well known. In the case of infinite-horizon linear-quadratic NZS differential games, the Nash policies are obtained by solving a set of coupled algebraic Riccati equations (AREs) \citep{LewisVraSyr2012}. Solving the set of coupled AREs is not trivial, and different algorithms to obtain such solutions have been proposed \citep{LiGaj1995,Engwerda2007,PossieriSass2015}. These algorithms require full knowledge of the mathematical model of the system.

In recent years, there has been an increasing interest in control design using only measured data. Many of the recent results in data-based control are based on Willems' fundamental lemma \citep{WillemsRapMarDe2005}, which states that every input-output trajectory of a dynamical linear system can be expressed in terms of a single persistently excited trajectory. In the current literature, there are now many different data-based control methods for discrete-time (see, e.g., \cite{MarkovskyDor2023} and the references therein) and continuous-time systems \citep{RapisardavanCam2024,EisingCor2025,LopezMulAUT2026}. Some model-free solutions of nonzero-sum differential games have been obtained using dynamic programming methods \citep{Vamvoudakis2015,Songetal2021,Xieetal2025}. These methods often exploit the properties of the Kronecker product, resulting in inefficient computations \citep{LopezMuCDC2023}. Data-based solutions for discrete-time NZS games have also been obtained using ideas based on Willems' lemma in \cite{Nortmannetal2024}, but those algorithms require collecting new data from the system at every iteration.

The literature discussed above concerns results for deterministic NZS games, where noise-free measurements of the full state vector are assumed to be available. In \cite{RhodesLuen1969,SaksenaCruz1982}, the stochastic nonzero-sum differential game is solved, where each player has access to a different noisy output signal. As it is discussed there, the truly optimal (Nash equilibrium) strategies require the use of infinite-dimensional observers. A practical solution is obtained by determining instead the optimal controllers based on an $n$-dimensional observer. The solution of this problem does not currently have a data-based counterpart. Note that in \cite{ZhangWanCao2024}, a different stochastic differential game is solved.

The main contribution of this paper is leveraging the continuous-time Willems' lemma in \citep{LopezMuCDC2022} to obtain model-free solutions of NZS games. We show how to collect persistently excited data only once from the multiagent system to design the data-based controllers. Both the deterministic and the stochastic NZS games are formulated and solved. For the stochastic case, as it is standard in the literature on data-based state estimation \citep{WolffLoMuECC2022,TuranFer2022}, we assume that the state of the system can be measured in an offline phase, whereas only output measurements are available during the online operation. This allows to use input-state-output persistently excited data to obtain the desired results.

The remainder of this paper is structured as follows. Section \ref{secprel} presents preliminary results and definitions. The data-based solutions of NZS games are presented in Section \ref{secsol}, including the deterministic game (Section~\ref{subsecdet}) and the stochastic one (Section~\ref{subsecstoc}). Section \ref{secsim} presents a simulation example and Section \ref{seccon} concludes the paper.

\section{Preliminaries}
\label{secprel}

\subsection{Deterministic nonzero-sum differential games}
\label{subsecpreldet}

Consider a dynamical system of the form
\begin{equation}
	\dot x = A x + \sum_{i=1}^N B_i u_i
	\label{detgamesys}
\end{equation}
where each input $u_i \in \R^{m_i}$ is used to control the state $x \in \R^n$ while minimizing the cost function
\begin{equation}
	J_i = \int_0^\infty \left( x^\top Q_i x + u_i^\top R_{ii} u_i + \sum_{j \neq i} u_j^\top R_{ij} u_j \right) dt,
	\label{costi}
\end{equation}
with $Q_i \succeq 0$, $R_{ii} \succ 0$, and $R_{ij} \succeq 0$. This is the formulation of a nonzero-sum (NZS) game, where the input $u_i$ corresponds to the control action of player $i$. Since the state $x$ is affected by all the inputs $u_i$, each cost $J_i$ depends on the actions of all players. A Nash equilibrium is defined as a set of control policies $u_i^*$, $i=1,\ldots,N$, such that
\begin{equation*}
	J_i(u_i^*,u_{-i}^*) \leq J_i(u_i,u_{-i}^*)
\end{equation*}
for every input $u_i$ and every player $i$. Here, $u_{-i}$ denotes the set of inputs of all players except player $i$. In a NZS game there may be no Nash equilibrium solution, a unique solution, or multiple solutions \citep{Engwerda2005}.

When the model  \eqref{detgamesys} is known, a Nash equilibrium can be obtained by policies of the form $u_i^* = -K_i^* x$, where
\begin{equation}
	K_i^* = R_{ii}^{-1} B_i^\top P_i,
	\label{kstardef}
\end{equation}
and the matrices $P_i \succ 0$ solve the coupled algebraic Riccati equations (ARE) (\cite{LewisVraSyr2012})
\begin{multline}
	0 = P_i A + A^\top P_i + Q_i - P_i S_i P_i \\
	+ \sum_{j \neq i} \left( P_j B_j R_{jj}^{-1} R_{ij} R_{jj}^{-1} B_j^\top P_j - P_j S_j P_i - P_i S_j P_j \right),
	\label{care}
\end{multline}
for $i=1,\ldots,N$, where
\begin{equation}
	S_i = B_i R_{ii}^{-1} B_i^\top.
	\label{sdef}
\end{equation}
Diverse methods to solve the coupled AREs \eqref{care} have been studied in the literature \cite{LiGaj1995,Engwerda2007,PossieriSass2015}.

\subsection{Stochastic nonzero-sum differential games}
\label{subsecprelsto}

In \cite{RhodesLuen1969}, a stochastic zero-sum game was formulated and solved. This result was extended to nonzero-sum games in \cite{SaksenaCruz1982}. In these articles and many others (\cite{ChongAth1971,LiGaj1995}), the game is formulated as having only two players for notation simplicity. The extension to the multi-player case can be straightforwardly obtained. Thus, consider now the dynamical system 
\begin{subequations}
	\begin{align}
		\dot x &= A x + B_1 u_1 + B_2 u_2 + w \\
		y_i & = C_i x + v_i, \quad i=1, 2,
	\end{align}
	\label{stogamesys}
\end{subequations}%
where the pairs $(A,C_1)$ and $(A,C_2)$ are observable, $y_i \in \R^{p_i}$ is the output measurement of player $i$, and the noise vectors $w \in \R^n$, $v_1 \in \R^{p_1}$, and $v_2 \in \R^{p_2}$ are independent Gaussian disturbances with zero mean and covariances $W$, $V_1$, and $V_2$, respectively. The initial condition of the state is also a Gaussian random vector, independent of the noises $w$, $v_1$, and $v_2$, with mean $\bar x_0$ and covariance
\begin{equation*}
	E \{ x(0) x(0)^\top \} = M_0.
\end{equation*}

The objective of each player is to minimize the expected cost $E \{ J_i \}$ with $J_i$ as in \eqref{costi}. Since the players do not have state measurements available, state observers are designed. It is known that, if no restriction is imposed over them, Nash equilibrium can only be achieved by using infinite-dimensional observers (\cite{RhodesLuen1969}). To avoid this problem, the objective of this game is to determine the optimal controllers for all agents, when they are restricted to use $n$-dimensional observers. 

The solution of this game is given by the dynamic controller (\cite{SaksenaCruz1982})
\begin{equation}
	u_i = -R_{ii}^{-1} B_i^\top P_i \hat x^i,
	\label{stocont}
\end{equation}
\begin{equation}
	\dot{\hat{x}}^i = F_i \hat x^i + B_i u_i + L_i \left( y_i - C_i \hat x^i \right),
	\label{stoobs}
\end{equation}
where $\hat x^1(0) = \hat x^2(0) = \bar x_0$, $P_i$ solves the AREs \eqref{care},
\begin{equation}
	F_1 = A - S_2 P_2 \left( I + (M_{20}-M_{21})(M_{00}-M_{01})^{-1} \right),
	\label{f1def}
\end{equation}
\begin{equation}
	F_2 = A - S_1 P_1 \left( I + (M_{10}-M_{12})(M_{00}-M_{02})^{-1} \right).
	\label{f2def}
\end{equation}
Moreover,
\begin{equation}
	L_i = M_{ii}C_i^\top V_i^{-1},
	\label{ldef}
\end{equation}
$S_i$ is as in \eqref{sdef}, and the matrix $M$, partitioned as 
\begin{equation}
	M = \begin{bmatrix}
		M_{00} & M_{01} & M_{02} \\
		M_{10} & M_{11} & M_{12} \\
		M_{20} & M_{21} & M_{22}
	\end{bmatrix}
\end{equation}
with $M_{ij} \in \R^{n \times n}$, satisfies the differential equation
\begin{equation}
	\dot M = \bar A M + M \bar A^\top + \bar B \bar W \bar B^\top
	\label{mdifeq}
\end{equation}
with initial conditions $M_{ij}(0) = M_0 + \bar x_0 \bar x_0^\top$ for $i = j = 0$, and $M_{ij}(0) = M_0$ otherwise. Here, $\bar W = \blkdiag \{ W, V_1, V_2 \}$,
\begin{equation}
	\bar A = \begin{bmatrix}
		A-S_1 P_1 -S_2 P_2 & S_1 P_1 & S_2 P_2 \\
		A-F_1-S_2 P_2 & F_1-L_1C_1 & S_2 P_2 \\
		A-F_2-S_1 P_1 & S_1 P_1 & F_2-L_2C_2
	\end{bmatrix}
	\label{abardef}
\end{equation}
and
\begin{equation*}
	\bar B = \begin{bmatrix}
		-I & 0 & 0\\
		-I & L_1 & 0\\
		-I & 0 & L_2
	\end{bmatrix}.
\end{equation*}

\subsection{Persistence of excitation for continuous-time systems}
\label{subsecrep}

\cite{LopezMuCDC2022} showed a persistence of excitation condition for continuous-time systems. Here, these results are formulated for case of the multiagent system \eqref{detgamesys}.

Consider an integer $\eta \in \mathbb{N}$ and a positive scalar $T~\in~\mathbb{R}_+$. Suppose that the input signals $u_i$, $i=1,\ldots,N$, are applied to \eqref{detgamesys} and an input-state trajectory is collected in the time interval $[0, \eta T]$. Define the following matrices for $0 \leq t \leq T$
\begin{align}
	\mathcal{H}_{T}(x(t)) & := \left[ \begin{array}{cccc}
		x(t) & x(t+T) & \cdots & x(t+(\eta-1)T) \end{array} \right], \label{hankrow} \\
		\mathcal{H}_{T}(u_i(t)) & := \left[ \begin{array}{cccc}
			u_i(t) & u_i(t+T) & \cdots & u_i(t+(\eta-1)T) \end{array} \right], \nonumber
\end{align}
 $i=1,\ldots,N$. Notice that, from \eqref{detgamesys} and \eqref{hankrow}, we can write $\frac{d}{dt} \mathcal{H}_{T}(x(t)) = A \mathcal{H}_{T}(x(t)) + \sum_{i=1}^N B_i \mathcal{H}_{T}(u_i(t))$. Taking the integral on both sides of this expression, we get
\begin{multline}
	\mathcal{H}_{T}(x(T)) - \mathcal{H}_{T}(x(0)) \\
	= A  \int_0^T \mathcal{H}_{T}(x(\tau)) d\tau + \sum_{i=1}^N B_i \int_0^T \mathcal{H}_{T}(u_i(\tau)) d\tau.
	\label{prehankdyn}
\end{multline}
To simplify the notation, define the data matrices
\begin{equation}
	\tilde{\mathcal{H}}_{S}(x) := \mathcal{H}_{T}(x(T)) - \mathcal{H}_{T}(x(0)),
	\label{xmats}
\end{equation}
\begin{equation*}
	\mathcal{H}_{S}(x) := \int_0^T \mathcal{H}_{T}(x(\tau)) d\tau, \quad \mathcal{H}_{S}(u_i)  := \int_0^T \mathcal{H}_{T}(u_i(\tau)) d\tau,
\end{equation*}
\begin{equation*}
	\mathcal{H}_S(u) := \begin{bmatrix}
		\mathcal{H}_S(u_1)^\top & \cdots & \mathcal{H}_S(u_N)^\top
	\end{bmatrix}^\top.
\end{equation*}
We highlight again that, from \eqref{prehankdyn}, we can write
\begin{equation}
	\tilde{\mathcal{H}}_{S}(x) = A \mathcal{H}_{S}(x) + \sum_{i=1}^N B_i \mathcal{H}_{S}(u_i).
	\label{hankdyn}
\end{equation}

The matrices $\mathcal{H}_{S}(x)$, $\mathcal{H}_{S}(u_i)$ satisfy an important rank property when the data used to compute them satisfy a persistence of excitation condition. For the following definition, consider the concatenation of control inputs $u := [u_1^\top \; \cdots \; u_N^\top]^\top \in \R^m$, with $m= \sum_{i=1}^N m_i$.

\begin{defn}[Piecewise constant PE input]
	\label{defctpe}
	A piecewise constant persistently exciting (PCPE) input of order $L$ is defined as ${u(t + jT) = \mu_j}$, $0 \leq t < T$, $j=0,\ldots,\eta-1$, where the sequence $\{ \mu_j \}_{j=0}^{\eta-1}$, $\mu_j \in \mathbb{R}^m$, is such that
	\begin{equation*}
		\text{rank} \left( \left[ \begin{array}{cccc}
			\mu_0 & \mu_1 & \cdots & \mu_{\eta-L} \\
			\mu_1 & \mu_2 & \cdots & \mu_{\eta-L+1} \\
			\vdots & \vdots & \ddots & \vdots \\
			\mu_{L-1} & \mu_{L} & \cdots & \mu_{\eta-1}
		\end{array} \right] \right) = mL,
	\end{equation*}
	and where the constant $T$ does not correspond to a pathological sampling time \citep{LopezMuCDC2022}.
\end{defn}

\begin{lem}[\cite{LopezMuCDC2023}, Lemma 4]
	\label{lempe}
	Consider the system \eqref{detgamesys}, let the pair $(A,[B_1 \; \cdots \; B_N])$ be controllable, and let $u= [u_1^\top \; \cdots \; u_N^\top]^\top$ be a PCPE input of order $n+1$. Then,
	\begin{equation}
		\text{rank} \left( \left[ \begin{array}{c}
			\mathcal{H}_S(x) \\ \mathcal{H}_S(u)
		\end{array} \right] \right) = n + m.
		\label{pecond}
	\end{equation}
\end{lem}

Note that collecting PE data from a multiagent system implies a degree of cooperation among the players, since the global input $u$ must be persistently exciting.

\section{Data-based solution of NZS games}
\label{secsol}

This section first shows useful data-based expressions for multiagent systems. Then, we present our data-based solutions to the deterministic NZS game described in Section \ref{subsecpreldet}, as well as the stochastic game in Section \ref{subsecprelsto}.

\subsection{Useful data-based expressions}

The rank condition \eqref{pecond} allows obtaining the data-based representation of closed loop systems \citep{DePersisTes2020}. In particular, we show the following result.

\begin{lem}
	\label{lemclsys}
	Consider the system \eqref{detgamesys} and let the pair $(A,[B_1 \; \cdots \; B_N])$ be controllable. Suppose that $u = [u_1^\top \cdots u_N^\top]^\top$ is a PCPE input of order $n+1$ as in Definition~\ref{defctpe}. Then, for any set of matrices $K_i \in \R^{m_i \times n}$, $i=1,\ldots,N$, there exists a matrix $\Gamma$ such that
	\begin{equation}
		\tilde{\mathcal{H}}_S(x) \Gamma = A- \sum_{i=1}^N B_i K_i.
		\label{clabk}
	\end{equation}
\end{lem}
\begin{pf}
	Since the concatenation of inputs $u$ is PCPE of order $n+1$, by Lemma \ref{lempe} we have that there exist a matrix $\Gamma$ such that
	\begin{equation}
		\begin{bmatrix} \mathcal{H}_S(x) \\ \mathcal{H}_S(u) \end{bmatrix} \Gamma = \begin{bmatrix} I & -K_1^\top & \cdots & -K_N^\top \end{bmatrix}^\top
		\label{aux1}
	\end{equation}
	for any matrices $K_i$. Moreover, from \eqref{hankdyn} we have
	\begin{equation*}
		\tilde{\mathcal{H}}_S(x) \Gamma = \begin{bmatrix} A & B_1 & \cdots & B_N \end{bmatrix} \begin{bmatrix} \mathcal{H}_S(x) \\ \mathcal{H}_S(u) \end{bmatrix} \Gamma
	\end{equation*}
	which, together with \eqref{aux1}, implies \eqref{clabk}. $\square$
\end{pf}

The algorithm that is described in the next subsection to solve the coupled AREs \eqref{care} requires knowledge of initial stabilizing matrices $K_i$. Such stabilizing controllers can be obtained from data by exploiting Lemma \ref{lemclsys}. Here, we suggest to solve an auxiliary, global optimal control problem to determine such controllers, since they will be convenient later in this section. Notice that this auxiliary problem considers the global input $u := [u_1^\top \; \cdots \; u_N^\top]^\top$ and does not correspond to a game-theoretic problem.

Thus, define some auxiliary weighting matrices $\mathcal{Q} \in \R^{n \times n}$, $\mathcal{R} \in \R^{m \times m}$. The matrix $\mathcal{Q} \succeq 0$ can be chosen arbitrarily, but it will be useful to define $\mathcal{R} = \blkdiag_i \{ R_{ii} \}$. The objective now is to determine the solution of the LQR problem defined by the matrices $(A,[B_1 \; \cdots \; B_N], \mathcal{Q}, \mathcal{R})$. In \cite{LopezMulAUT2026}, the LQR problem is solved for continuous-time systems using only measured data. Here, we rewrite this algorithm using our multiagent system formulation. First, determine the matrix $\mathcal{P} \in \R^{n \times n}$ that solves the optimization problem
\begin{align}
	\underset{P}{\text{maximize}} \quad & \trace (P) \label{optproblqr} \\
	\text{s.t.} \qquad & P\succ 0, \quad \mathcal{L}(P) \succeq 0, \nonumber 
\end{align}
where
\begin{multline}
	\mathcal{L}(P) = \mathcal{H}_S(x)^\top \mathcal{Q} \mathcal{H}_S(x) + \mathcal{H}_S(u)^\top \mathcal{R} \mathcal{H}_S(u) \\
	+ \mathcal{H}_S(x)^\top P \tilde{\mathcal{H}}_S(x) + \tilde{\mathcal{H}}_S(x)^\top P \mathcal{H}_S(x).
	\label{lp}
\end{multline}
Then, the obtained matrix $\mathcal{P}$ is used to find $\Gamma^0$ such that
\begin{equation}
	\left[ \begin{array}{c} \mathcal{H}_S(x) \\ \mathcal{L}(\mathcal{P}) \end{array} \right] \Gamma^0 = \left[ \begin{array}{c} I \\ 0 \end{array} \right].
	\label{gammasol}
\end{equation}
Finally, the matrices $K_i^0$ that solve this auxiliary LQR problem are given by
\begin{equation}
	K_i^0 = -\mathcal{H}_S(u_i) \Gamma^0.
	\label{dbk0}
\end{equation}

\begin{lem}
	Let the conditions in Lemma \ref{lempe} hold and compute the matrix $K_i^0$ from the procedure described in \eqref{optproblqr}-\eqref{dbk0}. Then, $K_i^0$ is the optimal solution to the LQR problem defined by the matrices $(A,[B_1 \; \cdots \; B_N], \mathcal{Q}, \mathcal{R})$.
	\label{lemk0}
\end{lem} 
\begin{pf}
	The proof follows as in \cite[Theorem 17]{LopezMulAUT2026}, where it is shown that the global optimal controller has the form $K^0= -\mathcal{H}_S(u) \Gamma^0$, which implies \eqref{dbk0} for each player. $\square$
\end{pf}

Since the matrices $K_i^0$ are stabilizing, they will be used to initialize the algorithm in the next subsection. Besides, recall that these matrices satisfy $\begin{bmatrix} K_1^{0\top} & \cdots & K_N^{0\top} \end{bmatrix}^\top = \mathcal{R}^{-1} \begin{bmatrix} B_1 & \cdots & B_N \end{bmatrix}^\top \mathcal{P}$ and, therefore, 
\begin{equation}
	K_i^0 = R_{ii}^{-1} B_i^\top \mathcal{P}.
	\label{usefulk0}
\end{equation} 
In the following, it will be useful to determine data-based expressions for the matrices $S_i$ in \eqref{sdef} for all players. This can be achieved by exploiting \eqref{usefulk0} and Lemma~\ref{lemclsys}. 

First notice that, from \eqref{gammasol} and \eqref{dbk0}, $\Gamma^0$ satisfies
\begin{equation}
	\begin{bmatrix} \mathcal{H}_S(x) \\ \mathcal{H}_S(u) \end{bmatrix} \Gamma^0 = \begin{bmatrix} I & -K_1^{0\top} & \cdots & -K_N^{0\top} \end{bmatrix}^\top.
	\label{clgamma0}
\end{equation}
Moreover, let $\bar \Gamma_i$ satisfy \eqref{clgamma0} but replacing $K_i$ with the zero matrix, where $\bar \Gamma_i$ is guaranteed to exist due to the rank condition \eqref{pecond}. This means that, as in \eqref{clabk}, $\tilde{\mathcal{H}}_S(x) \Gamma^0 = A- \sum_{i=1}^N B_i K_i^0$ and $\tilde{\mathcal{H}}_S(x) \bar \Gamma_i = A- \sum_{j \neq i} B_j K_j^0$. Therefore, we have
\begin{equation*}
	\tilde{\mathcal{H}}_S(x) (\bar \Gamma_i - \Gamma^0) = B_i K_i^0 = B_i R_{ii}^{-1} B_i^\top \mathcal{P}.
\end{equation*}
Now, we can compute $S_i$ in \eqref{sdef} for all $i$ using data as
\begin{equation}
	S_i = \tilde{\mathcal{H}}_S(x) (\bar \Gamma_i - \Gamma^0) \mathcal{P}^{-1}.
	\label{dbsdef}
\end{equation}

\subsection{Data-based solution of the deterministic NZS game}
\label{subsecdet}

In \cite{LiGaj1995}, an iterative procedure to solve the coupled AREs \eqref{care} was proposed. Algorithm 1 below shows this method for the setting considered in this paper.
\begin{figure}[h]
	\hrule
	{\bf Algorithm 1: Model-based solution of coupled AREs (\cite{LiGaj1995})}
	{\hrule \small
		\begin{algorithmic}[1]
			\Procedure{}{}
			\State Let $k=0$ and initialize the matrices $K_i^0$ such that $A-\sum_{i=1}^N B_i K_i^0$ is Hurwitz.
			\State Solve the following $N$ decoupled Lyapunov equations for $P_i^k \succ 0$  
			\begin{multline}
				P_i^{k} \left( A-\sum_{i=1}^N B_i K_i^k \right) + \left( A-\sum_{i=1}^N B_i K_i^k \right)^\top P_i^{k} = \\
				- Q_i - (K_i^k)^\top R_{ii} K_i^k - \sum_{j \neq i} (K_j^k)^\top R_{ij} K_j^k
				\label{mblyap}
			\end{multline}
			\State Using the solutions $P_i^k$, update the feedback matrices
			\begin{equation}
				K_i^{k+1} = R_{ii}^{-1} B_i^\top P_i^{k}, \quad i=1,\ldots,N.
				\label{mbup}
			\end{equation}
			\State Set $k = k+1$ and go to step 3 until convergence. 
			\EndProcedure
			\hrule
		\end{algorithmic}
	}
\end{figure} 

Our data-based solution to this problem makes use of the following calculations. First, as in \eqref{clgamma0}, given some matrices $K_i^k$, $i=1,\ldots,N$, compute $\Gamma^k$ such that
\begin{equation}
	\begin{bmatrix} \mathcal{H}_S(x) \\ \mathcal{H}_S(u) \end{bmatrix} \Gamma^k = \begin{bmatrix} I & -K_1^{k\top} & \cdots & -K_N^{k\top} \end{bmatrix}^\top.
	\label{clgammak}
\end{equation}
Moreover, define $\Phi_i^k \in \R^{\eta \times \eta}$ such that
\begin{equation}
	\begin{bmatrix} \mathcal{H}_S(x) \\ \mathcal{H}_S(u_1) \\ \vdots \\ \mathcal{H}_S(u_i) \\ \vdots \\ \mathcal{H}_S(u_N) \end{bmatrix} \Phi_i^k = \begin{bmatrix} \mathcal{H}_S(x) \\ -K_1^k \mathcal{H}_S(x) \\ \vdots \\ -K_i^k \mathcal{H}_S(x) + \mathcal{H}_S(u_i) \\ \vdots \\ -K_N^k \mathcal{H}_S(x) \end{bmatrix},
	\label{phik}
\end{equation}
where, on the right-hand side, only the block row corresponding to the $i$th input includes a term of the form $\mathcal{H}_S(u_i)$. Such matrices $\Phi_i^k$ always exist due to \eqref{pecond}. From this definition and \eqref{hankdyn}, it is easy to show that
\begin{equation}
	\tilde{\mathcal{H}}_S(x) \Phi_i^k =  \left( A - \sum_{i=1}^N B_i K_i^k \right) \mathcal{H}_S(x) + B_i \mathcal{H}_S(u_i).
	\label{phidyn}
\end{equation}
Using $\Phi_i^k$, define also the operator $\bar \mL_i^k(P)$ as
\begin{multline}
	\bar \mL_i^k(P) := \mathcal{H}_S(x)^\top (\bar Q_i + P S_i P) \mathcal{H}_S(x) \\
	+ \mathcal{H}_S(u_i)^\top R_{ii} \mathcal{H}_S(u_i) + \mathcal{H}_S(x)^\top P \tilde{\mathcal{H}}_S(x) \Phi_i^k \\
	+ (\Phi_i^k)^\top \tilde{\mathcal{H}}_S(x)^\top P \mathcal{H}_S(x),
	\label{dblp}
\end{multline}
with $\bar Q_i = Q_i + (K_i^k)^\top R_{ii} K_i^k + \sum_{j \neq i} (K_j^k)^\top R_{ij} K_j^k$ and $S_i$ is as in \eqref{dbsdef}.

Our data-based solution to the deterministic NZS game is shown in Algorithm $2$. In Theorem~\ref{thdet}, we show the equivalence of the model-based Algorithm $1$ and the data-based Algorithm $2$.

\begin{figure}[h]
	\hrule
	{\bf Algorithm 2: Data-based solution of coupled AREs}
	{\hrule \small
		\begin{algorithmic}[1]
			\Procedure{}{}
			\State Collect data from system \eqref{detgamesys} by applying inputs $u_i$ such that $u = [u_1^\top \cdots u_N^\top]^\top$ is PCPE of order $n+1$.
			\State Let $k=0$, initialize the matrices $K_i^0$ by solving \eqref{optproblqr}-\eqref{dbk0}, and compute the matrices $S_i$ as in \eqref{dbsdef}.
			\State Using $K_i^k$, compute $\Gamma^k$ as in \eqref{clgammak} and $\Phi_i^k$ as in \eqref{phik}.
			\State Solve the following $N$ decoupled Lyapunov equations for $P_i^k \succ 0$  
			\begin{multline}
				P_i^{k} \tilde{\mathcal{H}}_S(x) \Gamma^k + (\Gamma^k)^\top \tilde{\mathcal{H}}_S(x)^\top P_i^{k} = \\
				- Q_i - (K_i^k)^\top R_{ii} K_i^k - \sum_{j \neq i} (K_j^k)^\top R_{ij} K_j^k.
				\label{dblyap}
			\end{multline}
			\State Using the solutions $P_i^k$, compute $\bar \Gamma_i^k$, $i=1,\ldots,N$, from
			\begin{equation}
				\begin{bmatrix} \mathcal{H}_S(x) \\ \bar \mL_i^k(P_i^{k}) \end{bmatrix} \bar \Gamma_i^k = \begin{bmatrix} I \\ 0 \end{bmatrix},
				\label{gammaik}
			\end{equation}
			with $\bar \mL_i^k(P_i^{k})$ as in \eqref{dblp}.
			\State Update the feedback matrices
			\begin{equation}
				K_i^{k+1} = - \mathcal{H}_S(u_i) \bar \Gamma_i^k, \quad i=1,\ldots,N.
				\label{dbup}
			\end{equation}
			\State Set $k = k+1$ and go to step 4 until convergence. 
			\EndProcedure
			\hrule
		\end{algorithmic}
	}
\end{figure}

\begin{thm}
	\label{thdet}
	Consider the system \eqref{detgamesys} and let the pair $(A,[B_1 \; \cdots \; B_N])$ be controllable. Then, Algorithm $2$ is equivalent to Algorithm $1$ in the sense that the same matrices $P_i^k$, $K_i^{k+1}$, $i=1,\ldots,N$, are obtained at every iteration.
\end{thm}
\begin{pf}
	The data-based version of \eqref{mblyap} in Algorithm $1$ is easy to obtain because, from Lemma \ref{lemclsys}, we have that $\tilde{\mathcal{H}}_S(x) \Gamma^k = A- \sum_{i=1}^N B_i K_i^k$. A direct substitution in \eqref{mblyap} yields \eqref{dblyap}. Hence, it only remains to be shown that \eqref{dbup} and \eqref{mbup} are also equivalent.
	
	The fact that $P_i^{k}$ solves \eqref{mblyap}, implies also that it satisfies
	\begin{equation}
		P_i^{k} A_c^k + (A_c^k)^\top P_i^{k} + \tilde Q_i - P_i^k B_i R_{ii}^{-1} B_i^\top P_i^k = 0,
		\label{auxriccati}
	\end{equation}
	where $A_c^k := A-\sum_{i=1}^N B_i K_i^k$ and $\tilde Q_i := \bar Q_i + P_i^k B_i R_{ii}^{-1} B_i^\top P_i^k$. Thus, $P_i^k$ can be interpreted as the solution of the ARE \eqref{auxriccati}, corresponding to an LQR problem defined by the matrices $(A_c^k,B_i,\tilde Q_i, R_{ii})$. The solution of this auxiliary LQR problem is given by $K_i^{aux} = R_{ii}^{-1} B_i^\top P_i^{k}$ which, since $P_i^k$ satisfies \eqref{mblyap}, corresponds precisely to \eqref{mbup} in Algoritm $1$. Such an LQR problem could be solved in a data-based fashion by using a similar procedure to \eqref{optproblqr}-\eqref{dbk0}, but with $\mL(P)$ replaced by $\bar \mL_i^k(P)$ (see the analysis in Section 4.2 of \cite{LopezMulAUT2026} for details about the data-based solution of the LQR problem). Since in our case the solution $P_i^k$ of \eqref{auxriccati} was obtained from the Lyapunov equation \eqref{dblyap}, it only remains to compute $K_i^{aux}$ similarly as in \eqref{gammasol}-\eqref{dbk0}, from which the expressions in \eqref{gammaik} and \eqref{dbup} are obtained. $\square$ 
\end{pf}

\subsection{Data-based solution of the stochastic NZS game}
\label{subsecstoc}

Similarly as in the previous subsection, here we obtain a data-based formulation of the existing solution of the stochastic NZS game described in Section \ref{subsecprelsto}. Again, this is achieved by collecting persistently excited data from the system \eqref{stogamesys}. However, in the following we assume that, during this (offline) data-collection phase, the players have access to noise-free measurements of the input-state-output of the system. Although this assumption is restrictive in general, it has been used in the data-based state estimation literature \citep{WolffLoMuECC2022,TuranFer2022}, and can be fulfilled if such offline measurements are obtained in a dedicated laboratory setting equipped with advanced sensors. Such sensors are not used during the online application of the controller/state estimator, when only noisy input-ouput information is available. Robustifying this method for the case of offline noisy data is a subject of future work.

Hence, recalling that this game is formulated for only two players for simplicity, assume that the pair $(A,[B_1 \quad B_2])$ is controllable. Then, after applying the inputs $u_1$, $u_2$ to the system \eqref{stogamesys} such that $u = [u_1^\top \quad u_2^\top]^\top$ is a PCPE input of order $n+1$, compute the data matrices $\tilde{\mathcal{H}}_S(x)$, $\mathcal{H}_S(x)$, and $\mathcal{H}_S(u_i)$, $i=1,2$, as in \eqref{xmats}, as well as the matrices
\begin{equation*}
	\mathcal{H}_{S}(y_i) := \int_0^T \mathcal{H}_{T}(y_i(\tau)) d\tau,	\quad i=1,2,
\end{equation*}
with $\mathcal{H}_{T}(y_i(t))$ defined similarly as \eqref{hankrow}. Notice that, since noise-free data is assumed to be collected, it holds that 
\begin{equation}
	\mathcal{H}_{S}(y_i) = C_i \mathcal{H}_{S}(x).
	\label{ycx}
\end{equation}

Now, the data-based version of the solution of the stochastic NZS game described in Section \ref{subsecprelsto} can be obtained as follows. First, notice that the feedback matrix in \eqref{stocont} for each player corresponds to the deterministic Nash equilibrium policy, which can be obtained from Algorithm~$2$. Moreover, substituting \eqref{stocont} in \eqref{stoobs}, the state observer can be written as
\begin{equation}
	\dot{\hat{x}}^i = \bar F_i \hat x^i + L_i \left( y_i - C_i \hat x^i \right),
	\label{stoobs2}
\end{equation}
where, taking $S_i$ as in \eqref{sdef},
\begin{equation*}
	\bar F_1 = A - S_1 P_1 - S_2 P_2 \left( I + (M_{20}-M_{21})(M_{00}-M_{01})^{-1} \right),
\end{equation*}
\begin{equation*}
	\bar F_2 = A - S_2 P_2 - S_1 P_1 \left( I + (M_{10}-M_{12})(M_{00}-M_{02})^{-1} \right).
\end{equation*}
Notice that we can also write these matrices as
\begin{equation*}
	\bar F_i = A -B_1 K_1^* -B_2 K_2^* 	- S_j P_j (M_{j0}-M_{ji})(M_{00}-M_{0i})^{-1},
\end{equation*}
where $i \neq j$, and with $K_i^*$ as in \eqref{kstardef}. Hence, we obtain a fully data-based method by finding model-free expressions for the matrices $\bar F_i$, $L_i$ and $\bar A$.

Obtaining $L_i$ is straightforward by noticing that
\begin{align*}
	C_i & = C_i \mathcal{H}_{S}(x) \mathcal{H}_{S}(x)^\top (\mathcal{H}_{S}(x) \mathcal{H}_{S}(x)^\top)^{-1} \\
	& = \mathcal{H}_{S}(y_i) \mathcal{H}_{S}(x)^\top (\mathcal{H}_{S}(x) \mathcal{H}_{S}(x)^\top)^{-1}.
\end{align*}
Hence, from \eqref{ldef} we have
\begin{equation*}
	L_i = M_{ii} (\mathcal{H}_{S}(x) \mathcal{H}_{S}(x)^\top)^{-1} \mathcal{H}_{S}(x) \mathcal{H}_{S}(y_i)^\top W_i^{-1}.
\end{equation*}

Now, using the matrices $K_i^*$ that result from Algorithm 2, and exploiting again the rank condition \eqref{pecond}, compute $\Gamma^*$ from the set of equations
\begin{equation}
	\begin{bmatrix} \mathcal{H}_S(x) \\ \mathcal{H}_S(u_1) \\ \mathcal{H}_S(u_2) \end{bmatrix} \Gamma^* = \begin{bmatrix} I \\ -K_1^* \\ -K_2^* \end{bmatrix}.
	\label{clgammastar}
\end{equation}
As in Lemma \eqref{lemclsys}, this matrix gives us a data-based representation of the closed loop system $\tilde{\mathcal{H}}_S(x) \Gamma^* = A- B_1 K_1^* - B_2 K_2^*$. Using this result, we can write
\begin{equation*}
	\bar F_1 = \tilde{\mathcal{H}}_S(x) \Gamma^* -S_2 P_2^* (M_{20}-M_{21})(M_{00}-M_{01})^{-1},
\end{equation*}
\begin{equation*}
	\bar F_2 = \tilde{\mathcal{H}}_S(x) \Gamma^* -S_1 P_1^* (M_{10}-M_{12})(M_{00}-M_{02})^{-1}.
\end{equation*}

Finally, the matrix $\bar A$ in \eqref{abardef} can be obtained as
\begin{equation}
	\bar A = \begin{bmatrix}
		\tilde{\mathcal{H}}_S(x) \Gamma^* & S_1 P_1 & S_2 P_2 \\
		S_2P_2 \tilde M_1 & \bar F_1+S_1P_1-L_1C_1 & S_2 P_2 \\
		S_1P_1 \tilde M_2 & S_1 P_1 & \bar F_2+S_2P_2-L_2C_2
	\end{bmatrix},
\end{equation}
where $\tilde M_1 = (M_{20}-M_{21})(M_{00}-M_{01})^{-1}$ and $\tilde M_2 = (M_{10}-M_{12})(M_{00}-M_{02})^{-1}$. This can be seen by substituting \eqref{f1def}-\eqref{f2def} in the first column of \eqref{abardef}, and performing the resulting cancellations. Moreover, in the remaining columns of \eqref{abardef} we used the fact that $F_i = \bar F_i + S_iP_i$.

\begin{rem}
	In this subsection, it was assumed that noise-free data could be collected from the system during the offline phase. Computing the matrices above using noisy data is akin to the presence of parametric uncertainties in the model-based case. See, e.g., \citep[Section IV]{LopezAlMuTAC2023} where convergence of an iterative algorithm with noisy data is analyzed. Notice that arbitrarily small magnitudes of the noise would imply that the obtained controllers are arbitrarily close to the optimal ones. 
\end{rem}

\section{Simulation results}
\label{secsim}

In this section we verify the applicability of the proposed methods with a numerical simulation. Since both the deterministic and the stochastic NZS games studied in this paper make use of Algorithm $2$, here we focus only on simulating the stochastic case.

\begin{figure}
\begin{center}
\includegraphics[width=8.4cm]{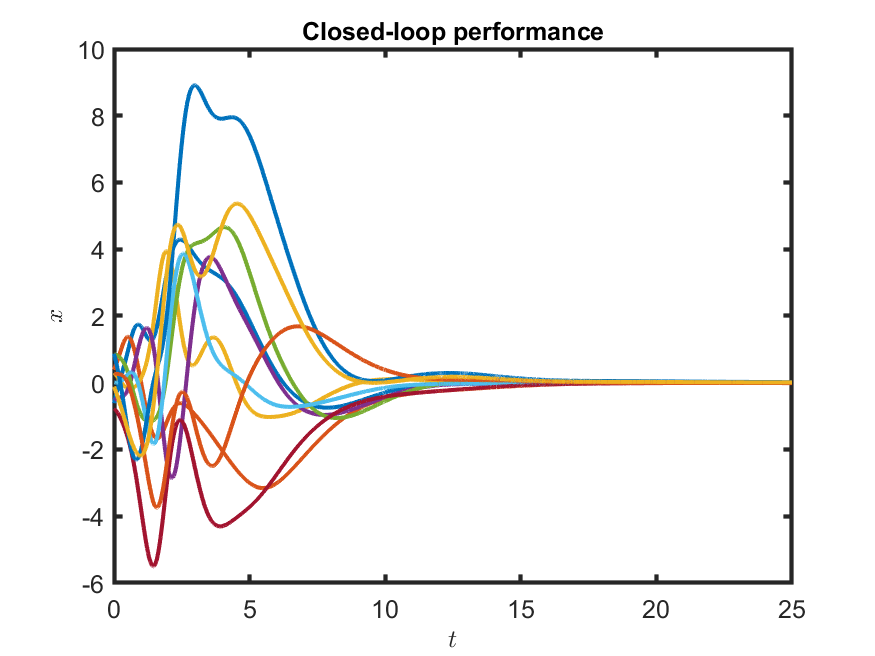}    
\caption{Performance of the proposed data-based controller.} 
\label{fig1}
\end{center}
\end{figure}

\begin{figure}
	\begin{center}
		\includegraphics[width=8.4cm]{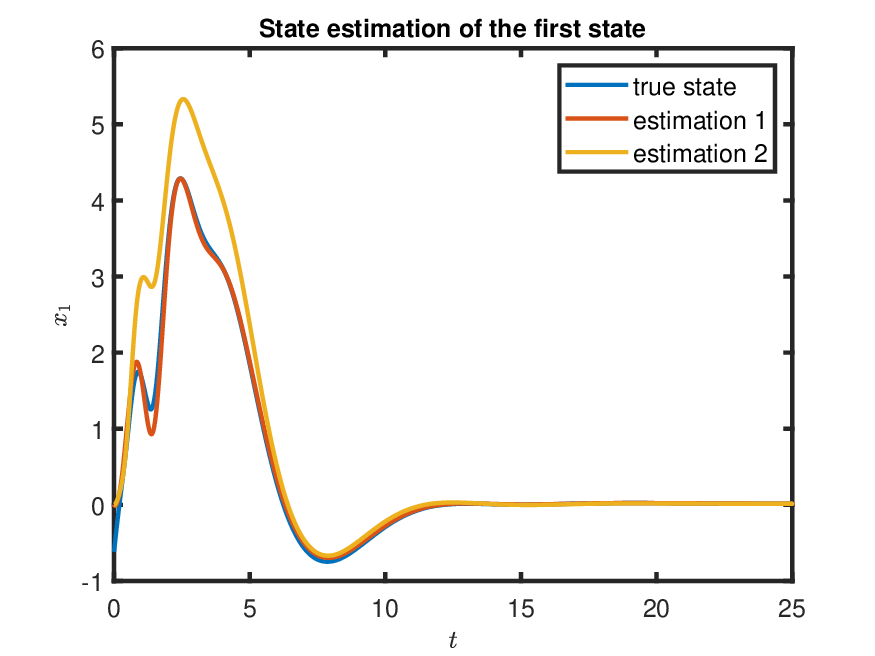}    
		\caption{State estimation of $x_1$ for both players} 
		\label{fig2}
	\end{center}
\end{figure}

We consider a system in the form \eqref{stogamesys} where the matrices $A$, $B_1$ and $B_2$, as well as the cost function matrices $Q_1$, $Q_2$, $R_{11}$, $R_{12}$, $R_{21}$, and $R_{22}$, are the same used in Numerical Example 1 in \citep{LiGaj1995}. This is a $10$-dimensional system, where each player uses a $2$-dimensional input. For the output matrices we choose $C_1 = [I \quad 0]$ and $C_2 = [0 \quad I]$, such that each player measures different states. The noise signals are assumed to have covariances $W=10^{-5} I$, and $V_1=V_2=10^{-2} I$, where the identity matrices have the proper dimensions. The objective for each player is to stabilize the system states while minimizing its expected cost $E \{ J_i \}$. Algorithm $2$ is run for 15 iterations. It can be verified that the obtained matrices $P_i^*$ satisfy the coupled AREs \eqref{care}. Then, the data-based state observer designed in Section~\ref{subsecstoc} is implemented in the control loop. The results are displayed in Figure \ref{fig1}, which shows that the 10 states are stabilized. Figure \ref{fig2} shows the state estimation of the first state, for both players.

\section{Conclusion}
\label{seccon}

We proposed novel data-based solutions for two classes of nonzero-sum differential games. First, we solve the deterministic game where all players can measure the complete system state. Then, we solved the stochastic game in which the players have only access to noisy output measurements during the online application of the controller. The obtained solutions are equivalent to their model-based counterparts and provide the desired outcomes under the same conditions. However, the proposed methods avoid the need for model knowledge by means of the collection of persistently excited data. As future work, we aim to relax the requirement in Section \ref{subsecstoc} of collecting noise-free data in the offline phase. Moreover, we will investigate the design of data-based solutions to differential graphical games, where distributed controllers must be obtained.


{\small
\bibliography{ifacconf}             

\begin{thebibliography}{26}
\providecommand{\natexlab}[1]{#1}
\providecommand{\url}[1]{\texttt{#1}}
\providecommand{\urlprefix}{URL }
\expandafter\ifx\csname urlstyle\endcsname\relax
  \providecommand{\doi}[1]{doi:\discretionary{}{}{}#1}\else
  \providecommand{\doi}{doi:\discretionary{}{}{}\begingroup
  \urlstyle{rm}\Url}\fi

\bibitem[{Basar and Olsder(1999)}]{BasarOls1999}
Basar, T. and Olsder, G.J. (1999).
\newblock \emph{Dynamic Noncooperative Game Theory}.
\newblock PA: SIAM, Philadelphia, USA, 2 edition.

\bibitem[{Chong and Athans(1971)}]{ChongAth1971}
Chong, C.Y. and Athans, M. (1971).
\newblock On the stochastic control of linear systems with different
  information sets.
\newblock \emph{IEEE Transactions on Automatic Control}, 16(5), 423--430.

\bibitem[{De~Persis and Tesi(2020)}]{DePersisTes2020}
De~Persis, C. and Tesi, P. (2020).
\newblock Formulas for data-driven control: Stabilization, optimality, and
  robustness.
\newblock \emph{IEEE Transactions on Automatic Control}, 65(3), 909--924.

\bibitem[{Eising and Cortés(2025)}]{EisingCor2025}
Eising, J. and Cortés, J. (2025).
\newblock When sampling works in data-driven control: Informativity for
  stabilization in continuous time.
\newblock \emph{IEEE Transactions on Automatic Control}, 70(1), 565--572.

\bibitem[{Engwerda(2005)}]{Engwerda2005}
Engwerda, J. (2005).
\newblock \emph{LQ dynamic optimization and differential games}.
\newblock John Wiley \& Sons, West Sussex, England.

\bibitem[{Engwerda(2007)}]{Engwerda2007}
Engwerda, J. (2007).
\newblock Algorithms for computing {N}ash equilibria in deterministic {LQ}
  games.
\newblock \emph{Computational Management Science}, 4(2), 113–140.

\bibitem[{Lewis et~al.(2012)Lewis, Vrabie, and Syrmos}]{LewisVraSyr2012}
Lewis, F.L., Vrabie, D., and Syrmos, V.L. (2012).
\newblock \emph{Optimal Control}.
\newblock John Wiley \& Sons, New Jersey, USA, 3 edition.

\bibitem[{Li and Gajic(1995)}]{LiGaj1995}
Li, T.Y. and Gajic, Z. (1995).
\newblock Lyapunov iterations for solving coupled algebraic {R}iccati equations
  of {N}ash differential games and algebraic {R}iccati equations of zero-sum
  games.
\newblock In G.J. Olsder (ed.), \emph{New Trends in Dynamic Games and
  Applications}, 333--351. Birkh{\"a}user Boston, Boston, MA.

\bibitem[{Lopez et~al.(2023)Lopez, Alsalti, and Müller}]{LopezAlMuTAC2023}
Lopez, V.G., Alsalti, M., and Müller, M.A. (2023).
\newblock Efficient off-policy {Q}-learning for data-based discrete-time {LQR}
  problems.
\newblock \emph{IEEE Transactions on Automatic Control}, 68(5), 2922--2933.

\bibitem[{Lopez and Müller(2022)}]{LopezMuCDC2022}
Lopez, V.G. and Müller, M.A. (2022).
\newblock On a continuous-time version of {W}illems' lemma.
\newblock In \emph{2022 IEEE 61st Conference on Decision and Control (CDC)},
  2759--2764.

\bibitem[{Lopez and Müller(2023)}]{LopezMuCDC2023}
Lopez, V.G. and Müller, M.A. (2023).
\newblock An efficient off-policy reinforcement learning algorithm for the
  continuous-time {LQR} problem.
\newblock In \emph{2023 62nd IEEE Conference on Decision and Control (CDC)},
  13--19.

\bibitem[{Lopez and Müller(2026)}]{LopezMulAUT2026}
Lopez, V.G. and Müller, M.A. (2026).
\newblock Data-based control of continuous-time linear systems with performance
  specifications.
\newblock \emph{Automatica}, 189, 113002.

\bibitem[{Markovsky and Dörfler(2023)}]{MarkovskyDor2023}
Markovsky, I. and Dörfler, F. (2023).
\newblock Identifiability in the behavioral setting.
\newblock \emph{IEEE Transactions on Automatic Control}, 68(3), 1667--1677.

\bibitem[{Nortmann et~al.(2024)Nortmann, Monti, Sassano, and
  Mylvaganam}]{Nortmannetal2024}
Nortmann, B., Monti, A., Sassano, M., and Mylvaganam, T. (2024).
\newblock Nash equilibria for linear quadratic discrete-time dynamic games via
  iterative and data-driven algorithms.
\newblock \emph{IEEE Transactions on Automatic Control}, 69(10), 6561--6575.

\bibitem[{Possieri and Sassano(2015)}]{PossieriSass2015}
Possieri, C. and Sassano, M. (2015).
\newblock An algebraic geometry approach for the computation of all linear
  feedback {N}ash equilibria in {LQ} differential games.
\newblock In \emph{2015 54th IEEE Conference on Decision and Control (CDC)},
  5197--5202.

\bibitem[{Rapisarda et~al.(2024)Rapisarda, van Waarde, and
  Çamlibel}]{RapisardavanCam2024}
Rapisarda, P., van Waarde, H.J., and Çamlibel, M. (2024).
\newblock Orthogonal polynomial bases for data-driven analysis and control of
  continuous-time systems.
\newblock \emph{IEEE Transactions on Automatic Control}, 69(7), 4307--4319.

\bibitem[{Rhodes and Luenberger(1969)}]{RhodesLuen1969}
Rhodes, I. and Luenberger, D. (1969).
\newblock Stochastic differential games with constrained state estimators.
\newblock \emph{IEEE Transactions on Automatic Control}, 14(5), 476--481.

\bibitem[{Saksena and Cruz(1982)}]{SaksenaCruz1982}
Saksena, V. and Cruz, J. (1982).
\newblock A multimodel approach to stochastic {N}ash games.
\newblock \emph{Automatica}, 18(3), 295--305.

\bibitem[{Song et~al.(2021)Song, Wei, Zhang, and Lewis}]{Songetal2021}
Song, R., Wei, Q., Zhang, H., and Lewis, F.L. (2021).
\newblock Discrete-time non-zero-sum games with completely unknown dynamics.
\newblock \emph{IEEE Transactions on Cybernetics}, 51(6), 2929--2943.

\bibitem[{Starr and Ho(1969)}]{StarrHo1969}
Starr, A.W. and Ho, Y.C. (1969).
\newblock Nonzero-sum differential games.
\newblock \emph{Journal of Optimization Theory and Applications}, 3, 184--206.

\bibitem[{Turan and Ferrari-Trecate(2022)}]{TuranFer2022}
Turan, M.S. and Ferrari-Trecate, G. (2022).
\newblock Data-driven unknown-input observers and state estimation.
\newblock \emph{IEEE Control Systems Letters}, 6, 1424--1429.

\bibitem[{Vamvoudakis(2015)}]{Vamvoudakis2015}
Vamvoudakis, K.G. (2015).
\newblock Non-zero sum {N}ash {Q}-learning for unknown deterministic
  continuous-time linear systems.
\newblock \emph{Automatica}, 61, 274--281.

\bibitem[{Willems et~al.(2005)Willems, Rapisarda, Markovsky, and {De
  Moor}}]{WillemsRapMarDe2005}
Willems, J.C., Rapisarda, P., Markovsky, I., and {De Moor}, B.L. (2005).
\newblock A note on persistency of excitation.
\newblock \emph{Systems \& Control Letters}, 54(4), 325--329.

\bibitem[{Wolff et~al.(2022)Wolff, Lopez, and Müller}]{WolffLoMuECC2022}
Wolff, T.M., Lopez, V.G., and Müller, M.A. (2022).
\newblock Data-based moving horizon estimation for linear discrete-time
  systems.
\newblock In \emph{2022 European Control Conference (ECC)}, 1778--1783.

\bibitem[{Xie et~al.(2025)Xie, Lu, Deng, Sun, and Chen}]{Xieetal2025}
Xie, K., Lu, M., Deng, F., Sun, J., and Chen, J. (2025).
\newblock Data-driven dynamic output feedback {N}ash strategy for multi-player
  non-zero-sum games.
\newblock \emph{Journal of Systems Science and Complexity}, 38, 597--612.

\bibitem[{Zhang et~al.(2024)Zhang, Wang, and Cao}]{ZhangWanCao2024}
Zhang, B.Q., Wang, B.C., and Cao, Y. (2024).
\newblock An online {Q}-learning method for linear-quadratic nonzero-sum
  stochastic differential games with completely unknown dynamics.
\newblock \emph{Journal of Systems Science and Complexity}, 37, 1907–1922.

\end{thebibliography}
}                                                    
                                                   
\end{document}